\documentclass[twocolumn]{aastex6}
\usepackage{color}
\slugcomment{ revised version\today}

\begin{document}

\newcommand       \e        	[1]{\times10^{#1}}
\newcommand       \msun        	{$M_{\odot}$}
\newcommand       \lsun      	{$L_{\odot}$} 
\newcommand	     \mpc              {Mpc$^{-3}$}
\newcommand	     \cm             {cm$^{-3}$}
\newcommand	     \yr              {yr$^{-1}$}
\newcommand	     \myr              {$M_{\odot}$~yr$^{-1}$}
\newcommand       \mic        	 {$\mu$m}
\newcommand       \nun		{$\nu$}
\newcommand        \lya		{Ly$\alpha$}
\newcommand        \ha		{H$\alpha$}
\newcommand	     \zmin      	{$z_{min}$}
\newcommand      \zmax      	{$z_{max}$}
\newcommand 	      \nwats            {nW m$^{-2}$ sr$^{-1}$}
\newcommand      \gray       {$\gamma$-ray}
\newcommand      \yd      {$\widehat Y_d$}
\newcommand     \nut       {$\widetilde{\nu}$}
\newcommand{\sms}[1]{{\mbox{{\footnotesize #1}}}}


\title{IRON: A KEY ELEMENT FOR UNDERSTANDING THE ORIGIN \\AND EVOLUTION OF INTERSTELLAR DUST}

\author{Eli Dwek}
\affil{Observational Cosmology Lab., Code 665 \\ NASA Goddard Space Flight Center,
Greenbelt, MD 20771, \\ e-mail: eli.dwek@nasa.gov}

\begin{abstract}
The origin and depletion of iron differ from all other abundant refractory elements that make up the composition of the interstellar dust. Iron is primarily synthesized in Type~Ia supernovae (SNe~Ia) and in core collapse supernovae (CCSN), and is present in the outflows from AGB stars. Only the latter two are observed to be sources of interstellar dust, since searches for dust in SN~Ia have provided strong evidence for the absence of any significant mass of dust in their ejecta. 
Consequently, more than 65\% of the iron is injected  into the ISM in gaseous form.
Yet, ultraviolet and X-ray observations along many lines of sight in the ISM show that iron is severely depleted in the gas phase compared to expected solar abundances. The missing iron, comprising about 90\% of the total, is believed to be locked up in interstellar dust. This suggests that most of the missing iron must have precipitated from the ISM gas by cold accretion onto preexisting silicate, carbon, or composite grains. Iron is thus the only element that requires most of its growth to occur outside the traditional stellar condensation sources. This is a robust statement that does not depend on our evolving understanding of the dust destruction efficiency in the ISM.
Reconciling the physical, optical, and chemical properties of such composite grains with their many observational manifestations is a major challenge for understanding the nature and origin of interstellar dust.
\end{abstract}

\section{INTRODUCTION}
Interstellar dust condenses in quiescent stellar outflows during the asymptotic giant branch (AGB) phase of their evolution, or in the explosive ejecta of CCSNe. Its presence in the ISM is manifested primarily by the extinction of starlight, the diffuse infrared (IR) and millimeter emission, the  scattering of diffuse starlight, and the depletion of refractory elements in the ISM compared to their expected solar abundances \citep{savage96,draine03,jenkins09}. During and after its injection into the ISM the newly-formed dust is subject to various physical processes, including: grain destruction by thermal and kinetic sputtering and by vaporizing grain-grain collisions in SN-generated shocks, and grain growth by cold accretion.
The first models for the evolution of interstellar dust already identified the discrepancy between the formation rate of dust in the different stellar sources, and their more rapid destruction rate in the ISM \citep{dwek79,dwek80b}. Grain growth by accretion, or a reduction in the grain destruction rates were suggested as possible ways to account for the observed abundance of interstellar dust \citep{dwek79,dwek80b,dwek98,tielens98,zhukovska08a,calura10}. 

Further attempts to resolve this discrepancy have led to a recent revaluation of the destruction rates of dust in the ISM \citep{jones11,bocchio14,slavin15}. The study of \cite{slavin15} derived a silicate lifetime of $\sim 2-3$~Gyr, a $\sim 2-3$ fold increase compared to previous studies. The discrepancy between the formation and destruction rates of silicates is therefore still subjected to change. Carbon dust, may be more efficiently destroyed, requiring its growth and re-formation in the ISM \citep{jones13}.

In this paper we show that iron provides the most robust case for the need of the growth of grains by accretion in the ISM. 
We show that because of its unique origin, iron is mostly injected into the ISM in gaseous form. Yet depletion studies show that most of the iron is absent from the ISM gas, and must therefore be locked up in dust. This discrepancy is not affected by uncertainties in the dust destruction rates, 
and therefore provides conclusive proof that most of the iron must have been incorporated into dust by cold accretion in the dense ISM.

The paper is organized as follows. In Section~2 discuss the origin of iron which, unlike other refractory elements, is predominantly formed in SN~Ia. The basic equations for following the evolution of the refractory elements of Mg, Si, and Fe in the solar neighborhood are discussed in Section~3. In Section~4 we calculate the maximum amount of Mg, Si, and Fe that can be locked up in thermally-condensed grains.    Section~5 compares the observed depletion trends of these elements, illustrating the distinct growth history of Fe bearing grains. The results of the paper are briefly summarized in Section~6. 

 \section{THE ORIGIN OF IRON}

Iron is the most tightly bound nucleus in nature with a binding energy of 8.8~MeV per nucleon, and therefore the heaviest element that can be synthesized by the fusion of lighter nuclei in massive stars. It has four stable isotopes ($A$ = 54, 56, 57, and 58), with almost all ($\sim 92$\%) of its observed natural abundance in $^{56}$Fe. 
The iron that is injected into the interstellar medium (ISM) is made from material that underwent hydrostatic O and Si burning. Its  isotopic signature is determined by the abundance of available free particles ($n$, $p$, and $\alpha$) generated in the photodissociation of $^{28}$Si (silicon burning) and finalized in the subsequent explosion. It can be synthesized either directly as $^{56}$Fe or as its radioactive precursor $^{56}$Ni depending on the neutron excess ratio, $\eta \equiv (N-Z)/A$, in the expanding material. The production of $^{56}$Fe will be dominant when $\eta \approx 4/56 \approx 0.071$, whereas $^{56}$Ni will be the dominant specie at low values of $\eta$ \citep{clayton68,arnett96}. The photodissociation of $^{28}$Si  occurs at temperatures that are high enough so that $\eta \approx 0$, and $^{56}$Ni is synthesized by the assembly of $\alpha$ particles. The energy released in the radioactive chain $^{56}$Ni ($\tau_{1/2}=6.1$d) $\longrightarrow$ $^{56}$Co ($\tau_{1/2}=77.2$d)  $\longrightarrow \, ^{56}$Fe provides most of the energy that powers Type~Ia and CCSN light curves.

Iron is primarily produced in SN~Ia events, caused by the explosion of a carbon-oxygen white dwarf that either accreted matter from a companion red giant star, or that merged with a white dwarf companion. It is also produced at the endpoint of the quiescent evolution of massive stars that end their life as CCSN. 
A 25~\msun\ supernova (SN) produces only about 0.07-0.23~\msun\ of iron, depending on various parameters such as the initial stellar metallicity, the internal stellar rotation and mixing, and the characteristics of the piston (its location in mass and its energy) that simulates the explosion \citep{nomoto06, heger03, heger10,nomoto13}. SN~Ia produce significantly more iron, theoretically about 0.32-1.1~\msun, depending on the ignition sites, and the geometry and propagation of the nuclear flame \citep{travaglio04,travaglio05, seitenzahl13} and observationally clustered around 0.4 and 1-1.4 \msun, depending on the explosion mechanisms \citep{childress15}. 

 \section{THE CHEMICAL EVOLUTION OF THE SOLAR NEIGHBORHOOD}
We use the analytical model described in \cite{dwek11a} to follow the chemical evolution of the solar neighborhood, defined as the region contained within a $\sim 1$~kpc radius around the sun. The solar neighborhood has been the subject of intensive studies in the past, starting with the pioneering work of \cite{tinsley74} and more recent studies by \citep[][and references therein]{dwek79,dwek80b,dwek98,zhukovska08a,calura10}. Here we concentrate on the evolution of Fe, as well as the main other, non-carbonaceous, refractory elements Mg and Si using the observational constraints on the stellar and gas masses, and the star formation, CCSN, and SN~Ia rates. 
We adopt the Kroupa stellar initial mass function (IMF) in all the calculations \citep{kroupa01}.  
\subsection{Stellar and Interstellar Gas Mass and the Star Formation Rate}
\begin{figure}[htbp]
  \begin{center}
\includegraphics[width=3.0in]{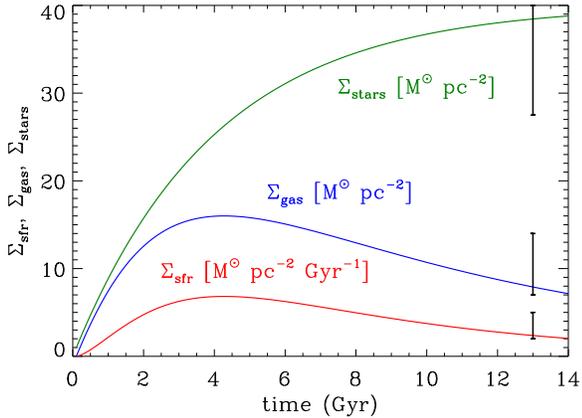}  
\end{center}
 \caption{{\footnotesize  The evolution of the surface densities of the star formation rate, and the mass of stars and the ISM  in the solar neighborhood. The vertical bars represent the observational constraints on their current values. Details in text. }\label{model}}
\end{figure}

The evolution of the gas mass surface density, $\Sigma_g$, in the instantaneous recycling approximation is given by:

\begin{equation}
\label{chemvol}
{d\Sigma_g(t)\over dt}= -(1-R_{ej})\, \Sigma_{sfr}(t) + \left({d\Sigma_g\over dt}\right)_{inf} \qquad .
\end{equation}
where $R_{ej}=0.50$ is the fraction of the initial stellar mass that is returned back to the ISM, calculated by averaging the mass-dependent ejecta over the stellar IMF, and 
 where the star formation rate (SFR) surface density, $\Sigma_{sfr}(t)$, is given by the Schmidt-Kennicutt relation \citep{kennicutt12}:
\begin{equation}
\label{sfr}
\Sigma_{sfr}(t) = \Sigma_{sfr}(t_0)\, \left[{\Sigma_g(t)\over \Sigma_g(t_0)}\right]^k \qquad,
\end{equation}
where $\Sigma_{sfr}(t_0)$ $\Sigma_g(t_0)$ are, respectively, the SFR and gas mass surface densities at the current epoch, $t_0=13$~Gyr, and where $k=1.5$. The metal-free infall rate was taken to be an exponential function:

\begin{equation}
\label{infall}
\left({d\Sigma_g\over dt}\right)_{inf} = \left({\Sigma_{inf}\over \tau}\right)\, \exp(-t/\tau) \qquad .
\end{equation}
 The stellar mass surface density at time $t$ is simply given by the difference:
\begin{equation}
\label{stars}
\Sigma_{\star}(t) = \Sigma_{inf}\, [1-exp(-t/\tau)] - \Sigma_g(t)  \qquad .
\end{equation}
In Equation~(\ref{infall}), $\Sigma_{inf}$ and $\tau$ are model parameters chosen to fit the current observational constraints on the surface densities of the current SFR, gas, and stellar mass in the solar neighborhood. 
The values of $\Sigma_{inf}$ and $\tau$ used in the fits were 40~\msun~pc$^{-2}$ and 4~Gyr, respectively.
Figure~\ref{model} presents the evolution of $\Sigma_g$, $\Sigma_{star}$, and $\Sigma_{sfr}$ in the solar neighborhood as as function of time.  The observational constraints  at the current epoch are indicated by vertical lines . 

 \subsection{The Evolution of Mg, Si, and Fe}
The Mg, Si, and Fe enrichment of the ISM is dominated by their synthesis in supernovae, with stellar winds primarily returning the mass of preexisting heavy elements back to the ISM. The Mg, Si, and Fe masses returned from AGB stars were calculated by multiplying the returned stellar masses, calculated for example by \cite{vandenhoek97}, by the solar abundances given by \cite{asplund09}. CCSNe yields, calculated for solar metallicities, were taken from \cite{heger10}, and scaled by a factor of 0.60 to take their dependence on metallicity into account.

SN~Ia yields depend on the location of the detonation point in the white dwarf interior.  \cite{seitenzahl13} calculated the nucleosynthesis yield of SNe~Ia using 3-dimensional models for a range of ignition geometries characterized by the density and symmetry of the ignition points. We adopt here the range of Mg, Si, and Fe yields from models N100, N100L and NH100H, characterized by 100 centralized ignition centers. The nominal Fe yield of this model is 0.64~\msun, with low and high values of 0.55 and 0.77~\msun. For comparison, an off-center model with fewer ignition centers, model N3, produced 1.1~\msun\ of iron. Observations of [Co~III]$\lambda$5893 emission during the nebular phase of the evolution of 7 SNe~Ia show that the masses of $^{56}$Ni synthesized in the ejecta cluster around 0.4 and 0.6-1.2~\msun, depending on the width of the SN light curve \citep{childress15}.
The range of Fe yields adopted here is therefore consistent with the observed range of Fe yields for a mixture of SN light curves.   

An additional factor determining the relative amount of Fe synthesized in core collapse and Type~Ia SNe is the relative rate of these events. 
The progenitors of SN~Ia are in the $\sim 2-8$~\msun\ mass range \citep{maoz08,greggio10}, and those of CCSN are stars more massive than 8~\msun\ \citep{smartt09,langer12}. Consequently, the relative contribution of the two sources depends on the stellar initial mass function (IMF). Because low-mass stars have a longer main sequence lifetime than their more massive counterparts, the relative contribution of SN~Ia and CCSN to the Fe abundance depends on the star formation history as well. Here we adopted the observational constraints of the relative SN~Ia and CCSN rates summarized by \cite{zhukovska08a} and \cite{calura10}. The nominal ratio of these rates varies from 0.15 to 0.25. 

\begin{figure}[tbp]
  \begin{center}
\includegraphics[width=3.0in]{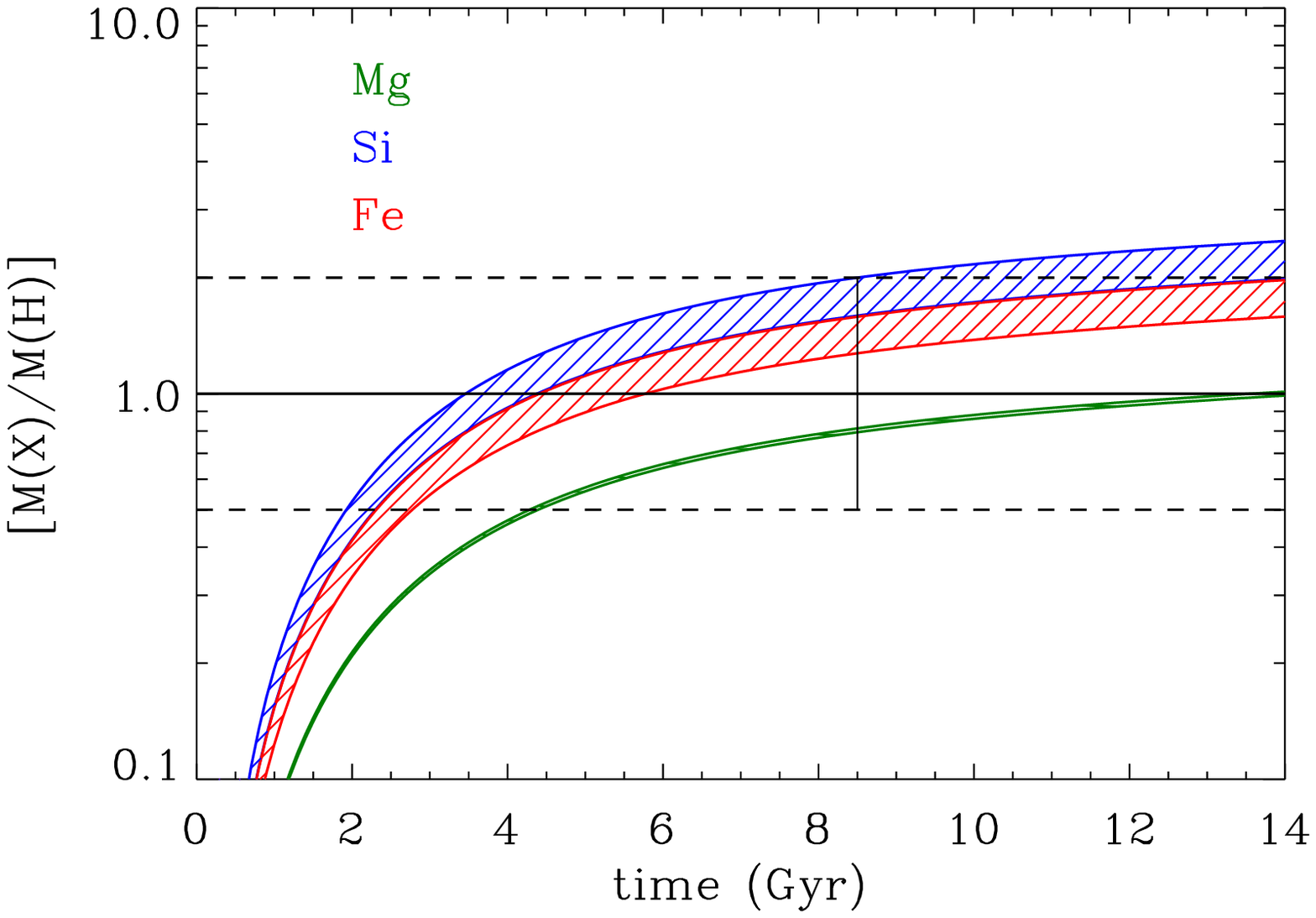}
\includegraphics[width=3.0in]{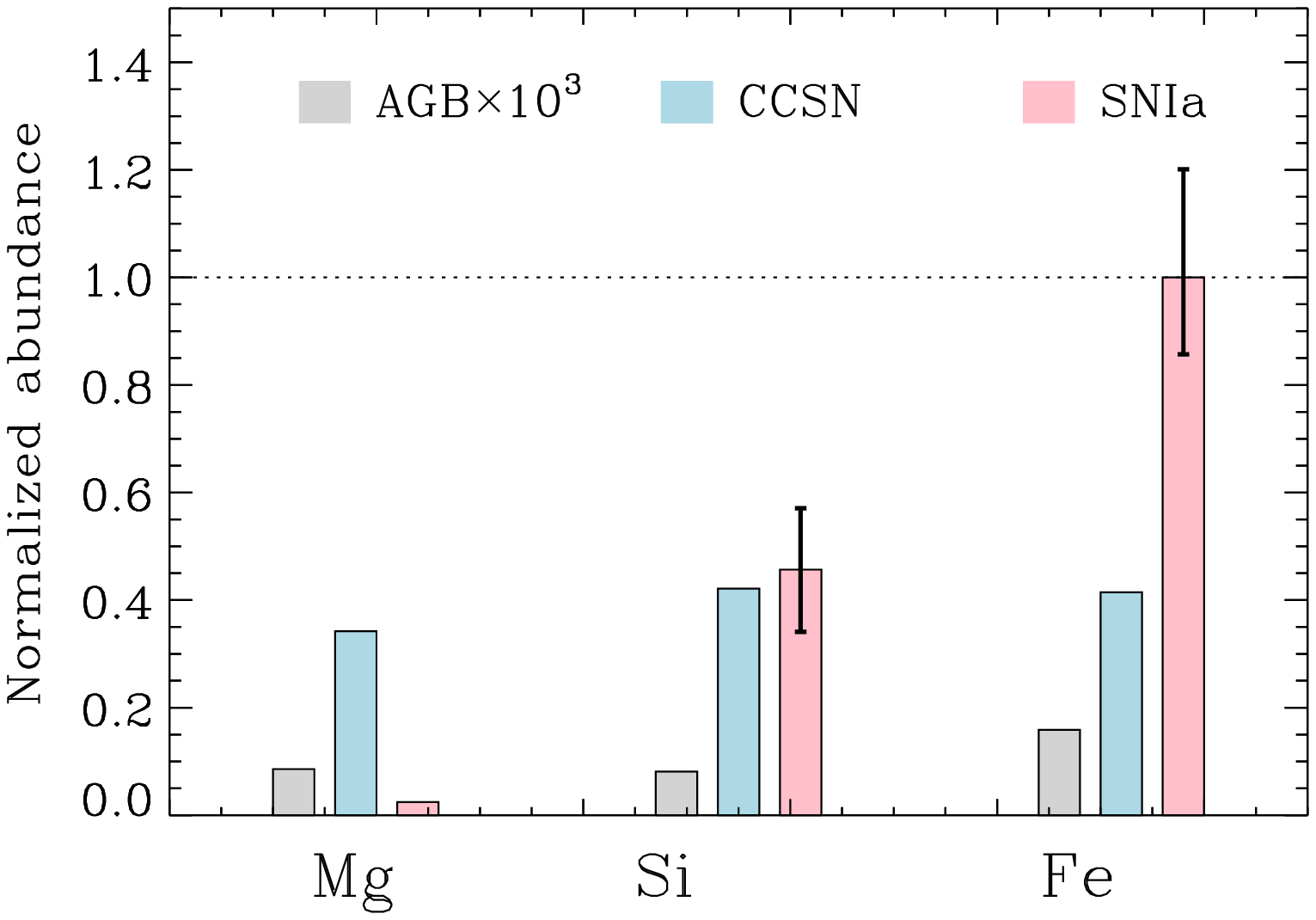}  
\end{center}
 \caption{{\footnotesize {\bf Top panel}: The evolution of Mg, Si, and Fe abundances, normalized to solar abundances, versus time. The horizontal dashed lines indicate deviations by a factor of two from solar abundances (thick horizontal line). The hatched area represents the range of nucleosynthetic yields of these elements in SN~Ia. The vertical solid line at 8.47~Gyr marks the age of the sun. {\bf Bottom panel}: The relative contribution of the different sources (AGB stars, CCSNe, and SN~Ia) to the total abundance of Mg, Si, and Fe, at the current epoch (13~Gyr), normalized to the SN~Ia contribution to the Fe abundance. The error bars represent the adopted range of Fe yields in SN~Ia.}  \label{abundances}}
 \end{figure}

Figure~\ref{abundances} (top panel) depicts the evolution of the mass of Mg, Si, and Fe, normalized to their solar values, as a function of time. A low value of 0.15 was adopted for the SN~Ia/CCSN ratio of the rates, which reproduces the solar abundances at the epoch of the Sun's birth at $t=8.47$~Gyr fairly well.
The bottom panel shows the contribution of the different stellar sources to the total abundances of these elements at the current epoch, normalized to the Fe contribution from SN~Ia. The figure illustrates the dominance of SN~Ia as the source of interstellar iron.

 \section{THE MAXIMUM FRACTION OF Mg, Si, and Fe IN THERMALLY-CONDENSED DUST}

 \subsection{Evidence for Dust in CCSNe Ejecta}
SN condensed dust can be observed either during its formation process, or later, by looking at the debris of the explosion before it has been completely mixed with the ISM. The most prominent example for dust formation in SNe is SN1987A, where $\sim 0.4$~\msun\ of dust has been inferred from the infrared (IR) emission detected about 20~yr after the explosion \cite{matsuura11, matsuura15}. About 0.1--0.2~\msun\ of SN condensed dust has been observed in the remnants of the Crab nebula \citep{gomez12a,temim12b,temim13}, and about $\sim 0.1$~\msun\ in that of Cassiopeia~A \citep{rho08,barlow10,arendt14}. A detailed review of searches for the evidence of dust formation in SNe has been presented by \cite{gall11c}.

The observed dust mass in CCSNe falls always short of the total mass of refractory elements present in their ejecta. In particular, most of the iron produced in CCSN is observed to be in the gas phase. A survey of the X-ray emitting ejecta of Cas~A gives an Fe mass of $\sim 0.09-0.13$~\msun\ in the hot gas \citep{hwang12}, comparable to the total mass of iron synthesized in the explosion. This value depends on the assumed volume filling factor of the emitting region, and should therefore be regarded as an upper limit. In any case, the very presence of iron in the X-ray emitting gas shows that not all of it condensed in the ejecta. In spite of these observations, we will assume that all the Fe produced in CCSNe condenses into dust.

 \subsection{Evidence for the Absence of Dust in SN~Ia Ejecta}

Type~Ia SN are less likely to form dust in their ejecta compared to CCSNe. The mass ejected in a SN~Ia event is typically $\sim 1.4$~\msun, significantly lower that the $\sim 5-15$~\msun\ of material ejected in a typical CCSN. With similar explosion energies and expansion velocities, the temperature and density of the SN~Ia ejecta decreases more rapidly compared to those in CCSNe ejecta, shortening the time available for the formation of molecules and  grains. Furthermore, the abundance and growth of the grains in SN~Ia ejecta is greatly impeded by the large abundance of radioactive $^{56}$Ni, which produces fast electrons and $\gamma$-rays that are effective at dissociating precursor molecules in this environment. Models for the formation of dust in SN~Ia show that any dust that forms will have small radii ($< 100$~\AA), and will probably be totally destroyed by the reverse shock which is generated as the ejecta is decelerated by the ambient ISM \citep{nozawa11}. The models produced no significant amount of iron dust. 

Extensive searches for newly-formed dust in the remnants of SN~Ia have been conducted with the {\it Spitzer} and {\it Herschel} satellites. Observations of remnants such as Kepler, \citep{blair07, williams12, gomez12b}, RCW~86 \citep{williams11b}, SN~1006 \citep{winkler13}, Tycho \citep{williams13, gomez12b}, and N103B \citep{williams14} have shown that all the IR emission from these remnants arises from shocked interstellar dust, with no evidence for any dust associated with the SN ejecta \citep{gomez12b}. A firm upper limit that is considerably less than 0.1~\msun\ was put on any dust present in the ejecta of Kepler's SNR \citep{blair07}.  For all practical purposes, we can safely conclude that all the Fe produced in SN~Ia is returned to the ISM in gaseous form.

 \subsection{Dust in Stellar Winds}

AGB stars are important sources of dust, forming carbonaceous dust when the C/O ratio in their ejecta exceeds unity, and forming silicate dust otherwise \citep{waters04}. For AGB stars with initial solar metallicity, silicates will form in stars with masses in the 1.0-1.7 and 5.0-8.0~\msun\ mass range \citep{nanni13}. Here we assume that all the Mg, Si, and Fe in the ejecta of stars in this mass range are locked up in dust. 

 \subsection{Maximal Fraction of Mg, Si, and Fe in Thermally-Condensed Dust}
In calculating the maximum possible depletion of Mg, Si, and Fe in thermally condensed dust, we will  assume that they condense with unit efficiency in the ejecta of CCSN and AGB stars, but that they are returned in gaseous form in the ejecta of SN~Ia.
This will provide a strong upper limit on the fraction of Fe that can be locked up in thermally condensed dust in the ISM.
 
\begin{figure}[tbp]
\includegraphics[width=3.5in]{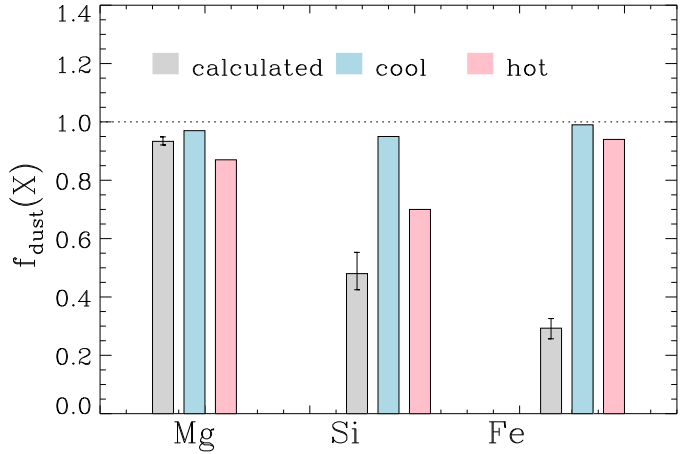}  
 \caption{{\footnotesize  The calculated fraction, $f_{dust}(X)$ of the total abundance of $X = $ \{Mg, Si, Fe\} that is  locked up in thermally-condensed dust (grey bars), is compared to observed values in the cool (blue) and hot (red) phases of the interstellar medium \citep{savage96}. \label{phases} }}
\end{figure}

The results of our calculations are presented in Figure~\ref{phases} which depicts the value of $f_{dust}(X)$, defined as the fraction of the abundance of the element $X=$\{Mg, Si, Fe\} that is locked up in thermally-condensed dust,
\begin{equation}
\label{fdust}
f_{dust}(X) \equiv X_{dust}/X_{tot} \qquad , 
\end{equation}
where $X_{dust}$ is the abundance of $X$ in the dust, and $X_{tot}$ is its total (gas+dust) abundance. A value of $f_{dust}(X)=1$ indicates that the element $X$ is totally absent from the gas phase of the ISM.  
The calculated value of $f_{dust}(X)$ represents the maximum attainable depletions of an element by thermal condensation, since it assumes a condensation efficiency of 100\% in AGB stars and CCSNe, and that none of the dust formed in CCSN is destroyed by the reverse shock generated by the propagation of the SN blast wave into its surrounding medium \citep{dwek05d,bianchi07,nozawa07,nath08,biscaro16,silvia10,silvia12,micelotta16}.

Comparison of the calculated values of  $f_{dust}(X)$ to the observed depletion fractions in the cool and warm phases of the ISM \citep{savage96} shows that, in general, they fall short of their depletion fractions in either of these phases of the ISM. The differences are not so large for Mg and Si bearing grains. The lifetime of silicate dust is still a subject of intense research, and may further increase when the clumpy nature of the ISM is taken into account \citep{dwek07b,jones11}. The observed depletions of Mg and Si may yet be explained by the efficiency of dust formation in the sources. 
The difference between the maximum attainable depletion and the observed depletions in the ISM is most pronounced for Fe. At most, only 35\% of the iron can be locked up in dust, compared to the observed 99 and 94\% depletions in, respectively, the cool and warm phases of the ISM.  

\begin{figure}[tbp]
\includegraphics[width=3.5in]{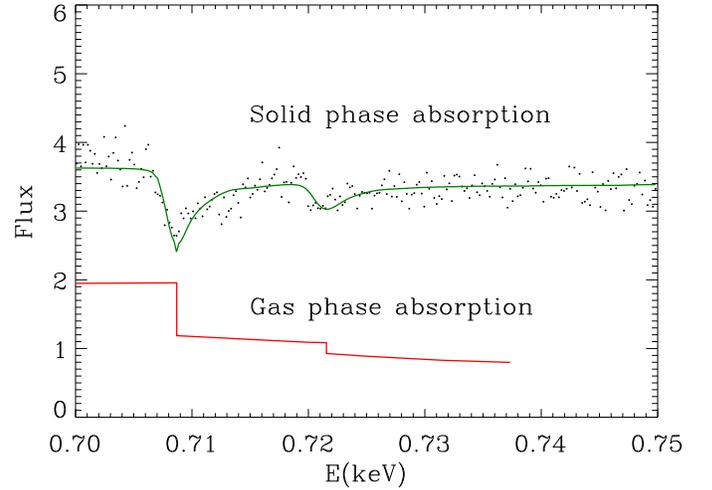}  
 \caption{{\footnotesize  Results of absorption line studies in the soft X-ray band region with the {\it Chandra} spacecraft (dots). The green solid line is a parametric fit using metallic iron data. The red line represents the iron absorption edges if all the iron was in the gas phase.  Data and the parametric fit were provided by T. Kallman, private communication}. \label{xray} }
\end{figure}

X-ray absorption line spectra provide an additional means for distinguishing between the gas and solid phase abundances of select refractory elements, and the composition of dust in the ISM \citep{lee09}. Indeed, X-ray absorption line studies towards eight X-ray binaries, obtained with the {\it Chandra} High Energy Transmission Grating Specrometer  provided evidence for the fact that most of the interstellar iron is locked up in dust \citep{gatuzz15}. Figure~\ref{xray} shows the observed soft X-ray spectrum around the location of the iron L2 and L3 absorption edges. The observations are represented by dots, and the green line represents a model in which the cross section and absorption edges were obtained from metallic iron data. The data could not be fitted with a pure iron gas model which is characterized by sharp absorption edges as shown by the scaled red line. The X-ray data provide therefore additional evidence that most of the interstellar iron is locked up in dust. So even if the Fe-bearing CCSN and AGB dust grains are not destroyed in the ISM, the very low $f_{dust}(Fe)$ value compared to the observed depletions provides the most compelling evidence for the need for its depletion in the ISM.

 \section{MAGNESIUM, SILICON, AND IRON DEPLETION TRENDS}

Additional insight into the nature of Fe-bearing dust can be obtained from a comparison between its depletion to that of other, non-carbonaceous, refractory elements. Carbon dust follows a distinct evolutionary cycle \citep{jones13}, and the analysis of its depletion pattern is outside the scope of this paper. The depletion of an element $X$ from the gas phase is usually defined as: $[X(gas)/H]\equiv \log\{N(X)/N(H)\}-\log(X/H)_{\odot} $, which expresses the reduction in its gas phase abundance (relative to hydrogen) compared to the expected value if all the $X$ atoms were in the gas phase. $N(X)$ and $N(H)$ are, respectively the observed abundance of the element $X$ and $H$ (in atomic plus molecular form) along the LOS. 

A detailed study of the depletion pattern of 17 different elements was  conducted by \cite{jenkins09}. Surveying over 243 lines of sight (LOS), he defined a parameter $F^{\star}$ to represents the overall strength of the depletion along any given LOS. A value of $F^{\star}=0$ was assigned to a LOS with the lowest overall depletion, whereas a value of $F^{\star}=1$ was assigned to the LOS toward the star $\zeta$~Oph, known for its strong depletions. Though most $F^{\star}$ values fall in the \{0,1\} interval, some LOS are characterized by lower or larger values.
The value of $F^{\star}$ reflects the history of grain growth and destruction in the gas along a given LOS.

  \begin{figure}[tbp]
\includegraphics[width=3.5in]{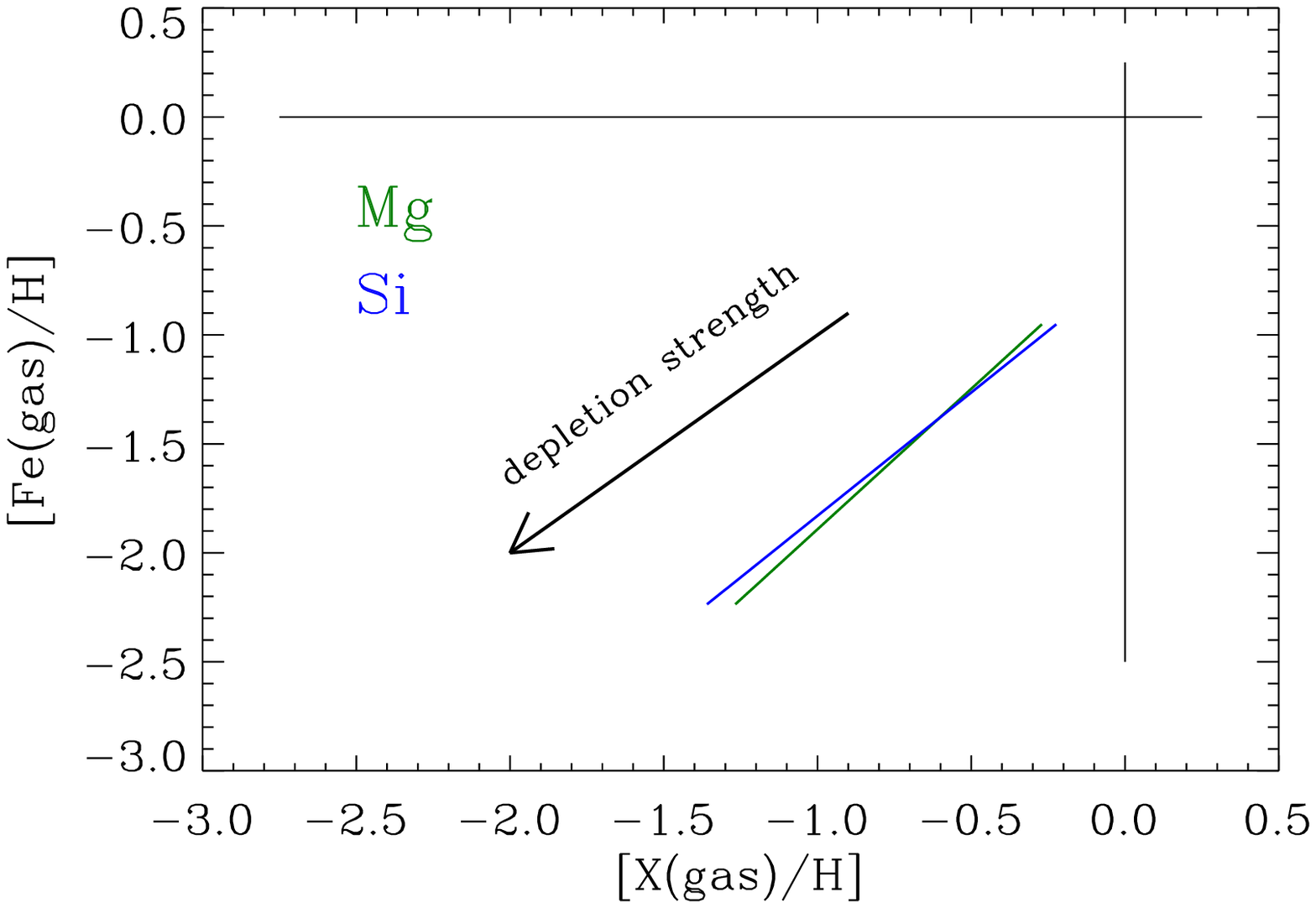}  
\includegraphics[width=3.5in]{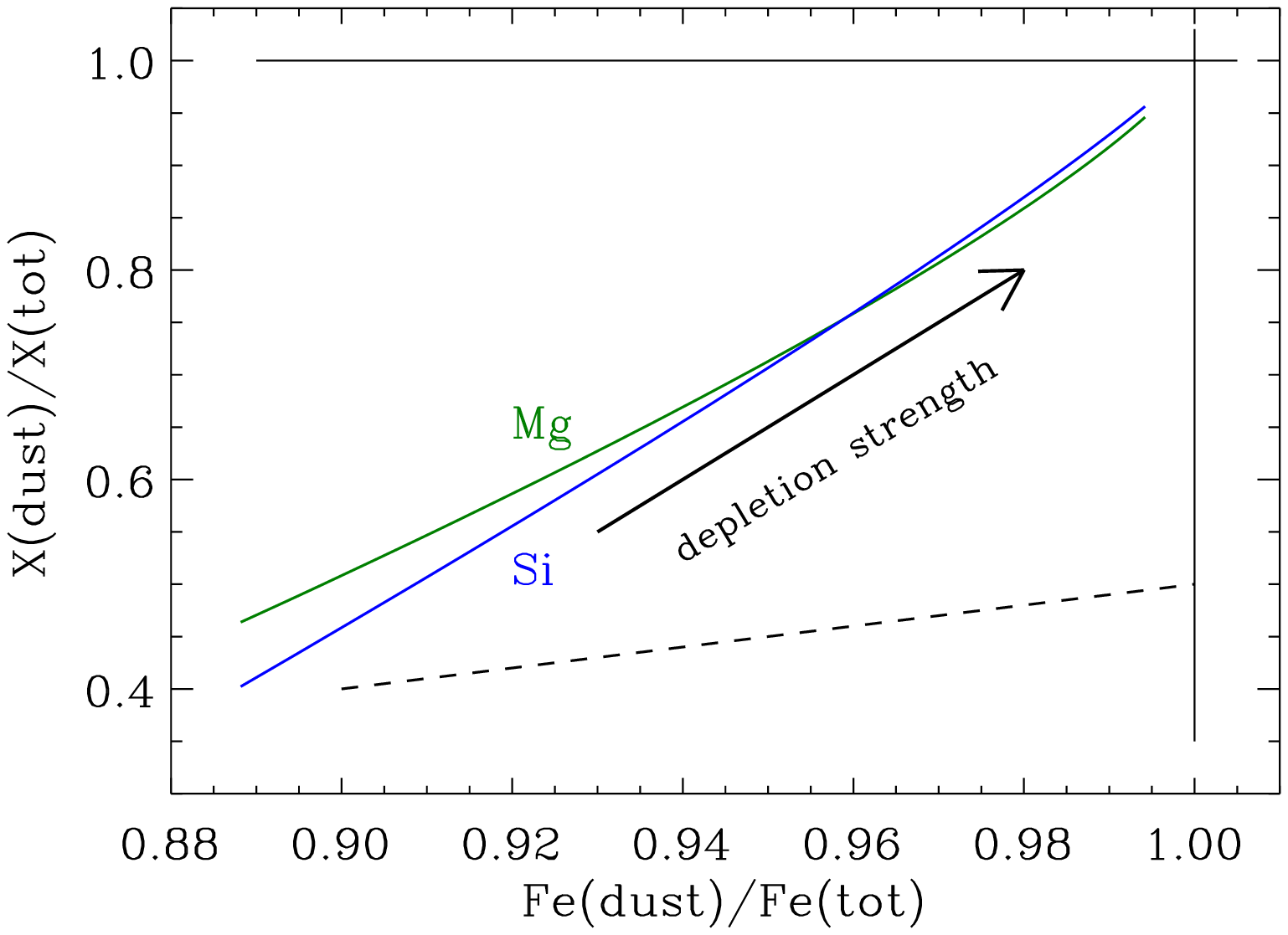} 
 \caption{{\footnotesize {\bf Top panel} Interstellar depletion factors  [Fe(gas)/H] versus [Mg(gas)/H] and [Si(gas)/H].  {\bf Bottom panel} The fraction of the total abundance of Mg and Si locked up in dust versus the Fe fraction locked up in dust. Curves were calculated using the elemental depletion parameters of \cite{jenkins09}. The dashed line shows the ISM depletion trend if Fe was depleted at the same rate or with the same stoichiometric ratio as Mg and Si.\label{depletion} }}
\end{figure} 

Figure~\ref{depletion} (top panel) plots the observed depletion of Fe versus that of Mg and Si.  
The bottom panel depicts the same information in the form of the fraction of the different elements locked up in the dust, given by 
$(X(dust)/H)/(X/H)_{\odot}=[1-10^{[X(gas)/H]}]$. The elements Mg, Si, and Fe attain the same maximum observed depletion, $[X(gas)/H]_1$ for $F^{\star}=1$ in the ISM, with about 95\% of the Mg and Si, and 99\% of the Fe locked up in dust. However, the minimal observed depletion, $[X(gas)/H]_0$ for $F^{\star}=0$, of Mg and Si is about $\sim 40-45$\%, whereas that of Fe  is $\sim 89$\%.  This means that any LOS in the ISM represents material in which the depletion of Fe has already grown from $\sim 35$ in the sources to $\sim 90$\% in the ISM.
Also shown in Figure~\ref{depletion} (bottom panel) is the correlation between the depletions of Mg, Si, and Fe if they were depleted to form minerals of the olivine group with a stochiometric ratio of 1:1:1 (dashed curve). The observed trends (colored lines) suggest that the progress of the depletion of Mg and Si is unrelated to that of Fe. If Mg, Si, and Fe accreted onto grains at this stochiometric ratio, then only $\sim 50$\% of the Mg and Si abundance could be locked up in dust when all the gas phase iron is exhausted. Mg and Si must therefore follow a distinct accretion process. Furthermore, the significantly larger observed value of $[Fe(gas)/H]_0$ compared to that of Mg and Si, in spite of its initial low value of $f_{th}(Fe)$, shows that Fe accretes more efficiently onto the pre-existing thermally-condensed grains than either Mg or Si.

Support for the different depletion trends is provided by recent spectral analysis of the composition of interstellar grains towards $\zeta$~Ophiuchi, a dense interstellar cloud representing the transition between the diffuse and dense phases of the ISM \citep{poteet15}. The observations show that Mg and Si are depleted in silicate compounds, whereas Fe is depleted in distinctly different solids. These observations provide important constraint on any model for an ISM origin for interstellar dust, namely preventing the accretion of iron onto silicate compounds. 

\section{SUMMARY}
Recent re-evaluation of dust lifetimes in the ISM concluded that the destruction rate of silicate dust may have been overestimated in the past, significantly narrowing the discrepancy between their formation rates in the different stellar sources, and their destruction rates in the ISM \citep{jones11, bocchio14, slavin15}. The need to reconstitute silicates by accretion in the ISM is therefore still a subject of further studies. 

In this paper we have shown that most of the iron depletion must take place in the ISM. This conclusion is robust, arising from the fact that most of the Fe is formed in SN~Ia, and therefore ejected into the ISM in gaseous form. The maximum percentage of Fe that can be depleted onto thermally-condensed dust is less  35\%, since we assumed a maximal condensation efficiency in AGB stars and CCSNe.  

The general issue of grain growth in the ISM invites a new set of well known problems that have to be reconciled with observational constraints \citep{jones11}. An additional problem is the ubiquitous presence of the 9.7 and 18~\mic\ silicate absorption features towards the Galactic Center \citep[][and references therein]{mccarthy80,fritz11,roche15}. The features are attributed to, respectively, the Si-O stretching and O-Si-O bending modes in silicates comprised of SiO$_4$ tetrahedra  \citep{khanna81}. Forming such tetrahedral structures requires the silicates to have formed or annealed at temperatures above $\sim 1000$~K for tens of hours \citep{hallenbeck00}. These conditions are unattainable if most of the silicates are grown by cold accretion in the ISM. Interstellar dust must therefore mostly comprise of composite refractory grains, carbon and silicates with organic refractory inclusions \citep{li01,zubko04}, or composite silicate carbon grains \citep{jones13}. Reconciling these requirements with the observed discrepancies between dust formation and destruction rates in the local ISM and galaxies remains a major problem in astrophysics.
 
Our work and observations by \cite{jenkins09} and \cite{poteet15} suggest that grains with accreted iron inclusions should be considered as a separate dust component of the ISM. Recent in-situ analysis of interstellar dust that penetrated the solar system supports this picture. The Cosmic Dust Analyzer on board the Cassini spacecraft revealed that the passing interstellar dust grains consist primarily of magnesium-rich grains of silicate oxide composition with iron inclusions \citep{altobelli16}. 

Iron inclusion in grains will have important effects on their alignment properties \citep{mathis86}.
Laboratory measurements of the optical properties of coated and composite dust particles, and exposure of composite grains containing metallic and organic material to UV and particle irradiation are important for furthering our understanding of the origin and nature of interstellar dust.   

\acknowledgements
In writing this paper I have benefitted from many enlightening conversation with Joe Nuth, and useful references provided by Steven Rodney. J.N. and Rick Arendt provided useful comments on an early version of the manuscript. I thank the referees Xander Tielens and Anthony Jones for their critical comments which led to improvements in the manuscript. This work was supported by NASA's 12-ADP12-0145 and 13-ADAP13-0094 research grants.

\clearpage
\bibliographystyle{$HOME/Library/texmf/tex/latex/misc/aastex52/aas.bst}
\bibliography{$HOME/Dropbox/science/00-Bib_Desk/Astro_BIB.bib}

 \end{document}